\documentstyle[]{article}
\font\cero=cmss10 scaled 1728 \font\uno=cmssbx10 scaled 1200
\begin{document}
\small{
\begin{flushleft}
{\cero  Canonical covariant formalism for Dirac-Nambu-Goto bosonic p-branes and the Gauss-Bonnet topological term
in string theory} \\[3em]
\end{flushleft}
{\sf Alberto Escalante}\\
{\it   Departamento de F\'{\i}sica, Centro de Investigaci\'on y de Estudios Avanzados del I.P.N., \\
Apdo Postal 14-740, 07000 M\'exico, D. F., M\'exico,  \\ 
Instituto de F\'{\i}sica, Universidad Aut\'onoma de Puebla,
Apartado postal J-48 72570, Puebla Pue., M\'exico}
(aescalante@fis.cinvestav.mx) \\[4em]
\noindent{\uno Abstract} \vspace{.5cm}\\
Using a covariant and gauge invariant geometric structure
constructed on the Witten covariant phase space for
Dirac-Nambu-Goto bosonic  p-branes propagating in a curved background, we
find the canonically conjugate variables, and the relevant
commutation relations are considered, as well as, we find the
canonical variables for the Gauss-Bonnet topological term in 
string theory.
\noindent \\

\begin{center}
{\uno I. INTRODUCTION}
\end{center}
\vspace{1em} \ Relativistic extended objects such as strings and
membranes, from the physical point of view can be considered as
fundamental building blocks of field theories, and represent the
more viable candidates for the quantum theory of gravity
\cite{1,2}. On the mathematical side, string and membranes have
been used to motivate unsuspected interplay among some
mathematical subjects. For example, at the perturbative level, it
is well known that string theory is related to the theory of
Riemann surfaces \cite{3} and some aspects of algebraic geometry
and
Mirror symmetry.\\
The quantization of such objects is a very complicated problem in
physics because, among other things, the theory is highly
non-lineal and the standard methods cannot be applied directly, even
for extended objects of simple topologies. However, in recent
works \cite{4,5}, using a covariant description of the canonical
formalism for quantization \cite{6}, and the deformations
formalism introduced by Capovilla-Guven [CG] in \cite{7}, the 
basis for the study of  the symmetries and the quantization aspects of
Dirac-Nambu-Goto bosonic p-branes [DNG] propagating in a curved background
space-time have been developed.  Such basis consist in the construction  a
covariant and gauge invariant symplectic structure on the
corresponding quotient phase space $Z$ (the space of solutions of
classical equations of motion divided by symmetry group volume)
instead of choosing a special coordinate system on the phase
space, with coordinates $p_{i}$ and $q^{i}$ as we usually find in
the literature.
The crucial observation of this work  is that such a choice is not necessary.\\
However, is remarkable to mention that recently there are works where using the constrained Hamiltonian approach \cite{5a} have been  keeping the reparametrization symmetry intact  \cite{5aa,  5b, 5c, 5d} ,  we will discuss this approach in the appendix part  with the results of this paper.\\
In this manner, the purpose of this article is to consider the
results obtained previously in \cite{4,5} and using basic ideas of
symplectic geometry in order to find; the canonically
conjugate variables, the corresponding  Poincar\'e charges, and
the fundamental Poisson brackets-commutators in a covariant
description, which is absent in the literature, as well as, to
find the canonical variables for Gauss-Bonnet term [GB] in string
theory
as first step to find a possible contribution due to such term at quantum level,
just as it was commented in \cite{39}.   \\
This paper is organized as follows. In Sect.II, using the results
presented in \cite{4,5} and standard ideas of symplectic geometry,
we identify the covariant canonically conjugate variables that we
will use in the development of this paper. In Sect.III, with the
results found in the last section, we obtain the  Poincar\'e
charges and its corresponding laws of conservation, confirming the
results presented in \cite{8}, as well as  the Poisson bracket and
the Poincar\'e algebra also are discussed. In Sect. IV using the
results given in \cite{39} we find the canonical variables for
[GB] topological term using an alternative method that we
introduced in previous  sections. In Sect. V we give some remarks and
prospects.\\

\setcounter{equation}{0} \label{c2}.

\noindent \textbf{II. Symplectic geometry in Dirac-Nambu-Goto bosonic p-branes}\\[1ex]
In \cite{4,5} a covariant and gauge invariant symplectic structure
for [DNG] bosonic p-branes propagating in a curved background has been
constructed, and is given by
\begin{equation}
\omega = \alpha \int_{\Sigma} \delta (-\sqrt{-\gamma}e{^{a}}_{\mu}
\delta X^{\mu} )d\Sigma_{a},
\end{equation}
where $e{^{a}}_{\mu}$ is a vector field tangent to the
world-volume created by the extendon, $\delta X^{\mu}$ is an
infinitesimal  space-time variation of the embedding \cite{7}, and
$\delta$, the deformation operator that acts as exterior
derivative on the phase space \cite{4,5}. It is important to
mention that $\omega$ turns out to be independent on the choice
of $\Sigma$,{\it i.e.}, $\omega_{\Sigma}=\omega_{\Sigma '}$ where
$\Sigma$ is a Cauchy p-surface, and it will be a very important
property of $\omega$, since it allows us to establish a connection
between functions and Hamiltonian vector fields on $Z$; this
subject will
be considered in the next paragraphs.\\
Because of $d \Sigma_{a}=\tau_{a}d \Sigma$, being $\tau_{a}$ a
normalized ($\tau_{a} \tau^{a}=-1$) timelike vector field tangent
to the world-volume, $\omega$ can be rewritten as
\begin{equation}
\omega= \int_{\Sigma} \delta \hat{p_{\mu}} \wedge \delta X^{\mu}d
\Sigma,
\end{equation}
where, $\hat{p_{\mu}}=- \sqrt{- \gamma}p_{\mu}$, and
$p_{\mu}=-\alpha e{^{a}}_{\mu} \tau_{a}$, satisfying
\begin{equation}
p_{\mu} p^{\mu}=- \alpha^{2},
\end{equation}
which is the mass shell condition for the p-brane. We can see that
$p_{\mu}$ is proportional to the timelike unit normal to $\Sigma$
into the world-volume.
\\
Since the formalism of deformations introduced in \cite{7} is
weakly covariant, the embedding functions depend of local
coordinates for the world-volume $(\xi^{a})$, and we have that
$\hat{p_{\mu}}=\hat{p_{\mu}}(\overrightarrow{\xi},\tau)$, and
$X^{\mu}=X^{\mu}(\overrightarrow{\xi},\tau)$, where we split the
local coordinates for the world-volume in an arbitrary evolution
parameter $\tau$ and coordinates $\overrightarrow{\xi}$ for
$\Sigma$ at fixed values of $\tau$. In this manner from equation
(2) we can identify  $\hat{p_{\mu}}(\overrightarrow{\xi},\tau)$ as
the canonical conjugate momentum to the embedding function
$X^{\mu}(\overrightarrow{\xi},\tau)$ for [DNG] p-branes in a
curved background. Therefore, any function $f$ on the covariant
phase space depends of $\hat{p_{\mu}}$ and $X^{\mu}$,
$f=f(\hat{p_{\mu}},X^{\mu})$.\\
Now, using basic ideas of symplectic geometry [9-11], we know that
if the symplectic structure $\omega$ of the theory under study is
invariant under a group of transformations $G$, (in our case
$\omega$ is invariant under space-time diffeomorphisms) which
corresponds to the gauge transformations of [DNG] theory
\cite{4,5}, the Lie derivative along a vector $V$ tangent to a
gauge orbit of $G$ of $\omega$ vanishes, that is,
\begin{equation}
\pounds_{V} \omega= V \rfloor \delta \omega + \delta (V \rfloor
\omega )= 0,
\end{equation}
where $\rfloor$ denotes the operation of contraction with $V$.
Since $\omega$ is an exact and in particular closed two-form,
$\delta \omega=0$ \cite{4,5}, and we have that Eq. (4), at least
locally, takes the form
\begin{equation}
V \rfloor \omega= -\delta H,
\end{equation}
where $H$ is a function on $Z$ which we call the generator of the
$G$ transformations \cite{6}. In this manner, the relation (5)
allows us to establish a connection between functions and
Hamiltonian vector fields on $Z$, just as we commented previously.\\
On the other hand, if $h$ and $g$ are functions on the phase
space, we can define using the symplectic structure $\omega$ a new
function $[f,g]$, the Poisson bracket of $h$ and $g$, as
\begin{equation}
[h,g]= V_{h} \rfloor g= - V_{g} \rfloor h,
\end{equation}
where $V_{h}$ and $V_{g}$ correspond to the Hamiltonian vector
fields generated by $h$, and $g$ respectively through Eq.
(4).\\
With these  results, in the next section we will calculate  in a
weakly covariant way the relevant Poisson bracket, the Poincar\'e
charges and their respective conservation laws; we show also the
closeness of the Poincar\'e
algebra.\\
\newline
\newline
\noindent \textbf{III. The Poisson bracket, Poincar\'e charges and Poincar\'e algebra in a flat space-time }\\[1ex]
For our aims, first we will find the vector fields associated with
the fundamental canonical variables on $Z$,
$(\hat{p_{\mu}},X^{\mu})$. For this, we use the expression (5),
finding  that
\begin{eqnarray}
 \!\!\ & X^{\alpha} & \!\!\ \quad \longrightarrow \quad V_{X^{\alpha}}=
- \frac{\partial}{\partial p_{\alpha}} \nonumber \\
 \!\! & \hat{p_{\alpha}}& \!\! \quad \longrightarrow \quad V_{p_{\alpha}}=
\frac{\partial}{\partial X^{\alpha}},
\end{eqnarray}
in this manner, using the equation (6) we have,
\begin{equation}
[X^{\mu}(\overrightarrow{\xi }',\tau),
X^{\nu}(\overrightarrow{\xi},\tau)] =
[\hat{p_{\mu}}(\overrightarrow{\xi
}',\tau),\hat{p_{\nu}}(\overrightarrow{\xi},\tau)]=0,
\end{equation}
\begin{equation}
[X^{\mu}(\overrightarrow{\xi},\tau),\hat{p_{\nu}}(\overrightarrow{\xi}',\tau)]
=  \delta{^{\mu}}_{\nu}
\delta(\overrightarrow{\xi}-\overrightarrow{\xi}'),
\end{equation}
where $\delta{^{\mu}}_{\nu}$ is the Kronecker symbol and
$\delta(\overrightarrow{\xi}-\overrightarrow{\xi}') $  the Dirac
delta function. In   Eqs. (8) and (9) we note that the
parameter $\tau$ is  a world-surface coordinate, this
implies a coordinate gauge fixing condition, because of the
invariance of the theory under world-surface parametrizations,
contrary to  the results found in \cite{9} that use a strongly
covariant
formalism.\\
On the other hand, if in Eq. (5) we choose $V= \epsilon^{\alpha}
\frac{\partial}{\partial X^{\alpha}}$, where $\epsilon^{\alpha}$
is a constant space-time vector, we find
\begin{equation}
V \rfloor \omega =- \delta[- \epsilon^{\mu} \tau_{a}(\alpha
\sqrt{- \gamma} e{^{a}}_{\mu})],
\end{equation}
where we can identify the linear momentum density
\begin{equation}
P^{a \mu}= -\alpha \sqrt{- \gamma} e^{a\mu},
\end{equation}
that is tangent to the world-volume and parallel to the tangent
vector $ e^{a\mu}$. Furthermore, using the Gauss-Weingarten
equations \cite{7}
\begin{eqnarray}
\nonumber \nabla_{a} e^{\mu}{_{b}} \!\! & = & \!\! -K{_{ab}}^{i}
n{_{i}}^{\mu}, \nonumber \\
\widetilde \nabla_{a} n{_{i}}^{\mu} \!\! & = & \!\! K_{abi}e^{\mu
b},
\end{eqnarray}
and remembering that the solutions to the  equations of motion for
[DNG] p-branes corresponds to extremal surfaces $(K^{i}=0)$ \cite
{5}, we find that the linear momentum are covariantly conserved,
this is
\begin{equation}
\nabla_{a}P^{a \mu}=0.
\end{equation}
On the other hand, we can express the total linear momentum
$P^{\mu}$ as
\begin{equation}
P^{\mu}= \int_{\Sigma} P^{a \mu} d \Sigma_{a} = \int_{\Sigma}
\hat{p^{\mu}} d \Sigma,
\end{equation}
where we can see that the total linear momentum, equation (14),
corresponds to the canonical momentum $\hat{p^{\mu}}$ given in
equation (2). Then the total lineal
momentum and the canonical momentum coincide.\\
Now, we will find the angular momentum of the extendon. For this
we rewrite our symplectic structure given in equation (1) as
\cite{5}
\begin{equation}
\omega= \int_{\Sigma} \delta  (\sqrt{- \gamma} e{^{a}}_{\mu}
\delta X^{\mu}) d \Sigma_{a}= \int_{\Sigma} \sqrt{- \gamma} n_{i
\alpha} \delta X^{\alpha} \wedge \widetilde \nabla^{a} (
n{^{i}}_{\mu} \delta X^{\mu})  d \Sigma_{a},
\end{equation}
where the $n{^{i}}_{\mu} $ are vector fields normal to the
world-volume. In this manner, for a vector field given by $V=
\frac{1}{2}[a{_{\alpha}}^{\beta} X^{\alpha}
\frac{\partial}{\partial X^{\beta}} - a{^{\alpha}}_ {\beta}
X^{\beta} \frac{\partial}{\partial X^{\alpha}}]$, with $a_{\alpha
\beta}=-a_{ \beta \alpha} $, the contraction $V \rfloor \omega $,
with $\omega$ expressed  as in Eq. (15), gives
\begin{eqnarray}
\nonumber V \rfloor \omega \!\!\ & = & \!\!  \sqrt{- \gamma}
\frac{1}{2} a_{\alpha \beta}   [n{_{i}}^{\beta} X^{\alpha}
\widetilde \nabla^{a} \phi^{i}- \phi_{i}n^{i \beta} e^{a \alpha}-
\phi_{i} X^{\alpha} \widetilde \nabla^{a} n^{i \beta} -
n{_{i}}^{\alpha} X^{\beta} \widetilde \nabla ^{a} \phi^{i} +
\phi_{i} X^{\beta} \widetilde \nabla
^{a}n^{i \alpha} + \phi_{i} e^{a \beta} n^{i \alpha}] \nonumber \\
\!\! & = & \!\!\ \sqrt{- \gamma} \frac{1}{2} a_{\alpha \beta}
[X^{\alpha} D_{\delta} e^{a \beta} - \phi_{i} n^{i \beta} e^{a
\alpha} - X^{\beta} D_{\delta} e^{a \alpha} + \phi_{i} e^{a \beta}
n^{i
\alpha} ]\nonumber \\
\!\!\ & = & \!\! \delta (a_{\alpha \beta} \frac{1}{2} [P^{a \beta}
X^{\alpha} - P^{a \alpha} X^{\beta} ]),
\end{eqnarray}
where we have used the equation (12) and $D_{\delta} e_{a}=
K{_{ab}}^{i} \phi_{i} e^{b} + \widetilde \nabla_{a} \phi_{i}
n^{i}$ \cite{7}; thus, from the last equation we can identify the
angular momentum of the p-brane given by
\begin{equation}
M^{a \beta \alpha}= \frac{1}{2} [P^{a \beta} X^{\alpha} - P^{a
\alpha} X^{\beta} ],
\end{equation}
and using equation (12) and the equation of motion $(K^{i}=0)$, we
get that
\begin{equation}
\nabla_{a}M^{a \beta \alpha }=0,
\end{equation}
this is, the angular momentum is covariantly conserved. It is
important to mention that the laws of conservation given by equations (13) and (18) correspond exactly to those found in
 \cite {8}, but obtained  in a different way.\\
 We define the total angular momentum  $M^{\alpha \beta}$  as
 \begin{equation}
 M^{\alpha \beta}= \int_{\Sigma} M^{\alpha \beta} d \Sigma_{a} =
 \int_{\Sigma}(\hat{p^{\beta}} X^{\alpha}- \hat{p^{\alpha}} X^{\beta}
 )d \Sigma.
 \end{equation}
In this manner, using equation (19) we can find the Hamiltonian
vector field associated to the angular momentum, using Eq. (5)
with $H^{\alpha \beta}=(\hat{p^{\beta}} X^{\alpha}-
\hat{p^{\alpha}} X^{\beta}) $ we find
\begin{equation}
V^{\beta \alpha} = \eta^{ \lambda \beta} \left( X^{\alpha }
\frac{\partial}{\partial X^{\lambda}} + \hat{p^{\alpha}}
\frac{\partial}{\partial \hat{p^{\lambda}}} \right)- \eta^{\lambda
\alpha} \left( X^{\beta } \frac{\partial}{\partial X^{\lambda}}+
\hat{p^{\beta}} \frac{\partial}{\partial \hat{p^{\lambda}}}
\right),
\end{equation}
where $\eta^{\mu \nu }$ is the Minkowski background metric. Thus,
using the equation (20) we can calculate the brackets $[M^{\mu
\nu}, M^{\alpha \beta}]$ and $[M^{\mu \nu}, P^{\alpha}]$, finding
\begin{eqnarray}
\nonumber [M^{\mu \nu}, M^{\alpha \beta}] \!\!\ & = & \!\!\
\eta^{\nu \alpha} M^{\mu \beta} + \eta^{\mu \alpha} M^{\beta \nu}
+ \eta^{\nu \beta}M^{\alpha \mu} + \eta^{\mu \beta} M^{\nu
\alpha},
\end{eqnarray}
\begin{eqnarray}
 [M^{\mu \nu},P^{\alpha} ] \!\!\ & = & \!\!\ \eta^{\mu \alpha} P^{\nu}-
\eta^{\alpha \nu} P^{\mu},
\end{eqnarray}
therefore, we can see that the Poincar\'e charges $P$ and $M$
indeed close correctly on the Poincar\'e algebra, such as in the
case of string theory using a standard canonical formalism
\cite{1}.\\
\newline
\newline
\noindent \textbf{IV. The canonical variables for The Gauss-Bonnet topological term in string theory }\\[1ex]
Along the same lines, in this section using the covariant
canonical formalism we will find the canonical variables for the
[GB] topological term in string theory. It is not possible to obtain this results usig the  conventional canonical formalism
because   we do not find any contribution to the equations of
motion, the reason being that  the field equations of the [GB] topological
term are proportional to the called Einstein tensor, and it does
not give any contribution to the dynamics in a two-dimension
worldsheet swept out by a string, since the Einstein tensor
vanishes for such a geometry \cite{39}. However, we identify from
a nontrivial covariant and gauge invariant symplectic structure
constructed in \cite{39} the canonical variables for such term.\\
It is difficult  to find the canonical variables for [GB]
topological term as in the  last section for [DNG] p-branes,  because
the symplectic structure for [GB] is not trivial \cite{39},
therefore, we need to use an alternative method  for such
purposes. This method consists in exploiting the frame gauge
dependence of the connection coefficients $\gamma{_{cd}}^{a}$
(which means that it can always be set equal to zero at any
single point chosen  by an appropriate choice of the relevant
frames) and the integral kernel of the geometric structure for
[GB] in string theory is given in terms of deformations of such
coefficients \cite{39}, and to rewrite the connection in terms of
the rotation (co)vector such as in \cite{17g}
which will be crucial for our developments. \\
For our aims, we need the covariant and gauge invariant symplectic
structure constructed in \cite{39} given by
\begin{equation}
\omega= \int_\Sigma \delta \Psi^{a}d\Sigma_{a} ,
\end{equation}
where $\Psi^{a}$ is identified as a symplectic potential for [GB]
in string theory
\begin{equation}
\Psi^{a}= \sqrt{ -\gamma} [\beta \gamma^{cd} \delta
\gamma{_{cd}}^{a}- \beta \gamma ^{ab} \delta \gamma{_{cb}}^{c}],
\end{equation}
here, $\beta$ is a constant, $\gamma{_{cd}}^{a}$ are the
connection coefficients \cite{7}, and $\delta$ is identified as exterior derivative on the phase space \cite{39}. \\
On the other hand, we know that for the case of string theory the
embedding 2-surface is characterized by an antisymmetric unit
tangent element tensor given by $\varepsilon^{\mu \nu}=
\varepsilon^{AB}l_{A}{^{\mu}}l{_{B}}^{{\nu}}$ (where
$\varepsilon^{AB} $ are the constant components of the standard
two-dimensional flat space alternating tensor) \cite{34}, thus, in
terms of the deformations formalism given in \cite{7} we can
introduce the rotation (co)vector
 $\rho_{a}$ given in terms of the frame gauge coefficients connection $\gamma_{ab}{^{c}}$ and $\varepsilon^{ab}$
 \begin{equation}
 \rho_{b}=\gamma_{bd}{^{c}}\varepsilon^{d}{_{c}},
 \end{equation}
 and this implies that
 \begin{equation}
  \gamma_{bd}{^{c}}=\frac{1}{2} \varepsilon^{c}{_{d}}
 \rho_{b},
 \end{equation}
 we can note that the frame gauge dependence of
 $\gamma_{ab}{^{c}}$ induces the same gauge dependence on $\rho_{a}
 $.\\
Considering  equations (24), (25) and the gauge dependence on
$\rho_{b}$, we can prove that $\Psi^{a}$ given by Eq. (23) takes
the form
\begin{equation}
\Psi^{a}=\sqrt{ -\gamma} \varepsilon^{ab} \rho_{b},
\end{equation}
thus, we can write the symplectic structure given in Eq. (22) as
\begin{equation}
\omega= \int_{\Sigma} \delta(\sqrt{ -\gamma} \varepsilon^{ab}
\tau_{a}) \wedge \delta(\rho_{b}) d\Sigma,
\end{equation}
this last equation has the same form of Eq. (2) for [DNG]
p-branes, in this manner we can identify  $p^{a}=\sqrt{ -\gamma}
\varepsilon^{ab} \tau_{a}$ and $q_{b}= \rho_{b}$ as canonical
variables for the [GB] topological term. It is remarkable to
mention that these canonical variables have indices of  the
worldsheet contrary to the canonical variables for [DNG] p-branes
that has indices of space-time (see Eq. (2)). This fact is
important because if we add the [GB] topological term to [DNG]
action in string theory and using the deformations formalism that
we used here to identify the canonical variables for this system
is very difficult, because combine variables wiht spacetime and
worldsheet indices.  However, this problem can be clarified using
the ideas presented in \cite{17g} where  a strongly
covariant formalism is used, and in this case we can find the canonical
variables for [GB] topological term with spacetime indices and
perhaps could be more easy to work the system [GB] and [DNG] in
string theory \cite{17b}.
\newline
\newline
\newline
\noindent \textbf{V. Conclusions and prospects}\\[1ex]
\newline
As we can see, identification of  the covariant canonical variables for
[DNG] p-branes is  easy because the expression of the
symplectic structure for the theory under study is simple. In this
manner, we could find the Poincar\'e charges, conservation laws in
a different way presented in \cite{8}, and construct the relevant
Poisson brackets. With these results we have the elements to
study the quantization aspects in a covariant way
and in  particular case of string theory which is absent in the literature. \\
It is important to emphasize that the canonical variables for [GB]
found in this paper has worldsheet indices, and if we add the [GB]
term in any action describing strings, for example [DNG] system,
we need to identify  the pullback on the canonical variables found
of  this paper to obtain the  canonical variables in terms of
spacetime indices, however, using the strongly covariant formalism
used in \cite{17g} this problem can be clarified. The only problem
 is to find a canonical transformation that
leaves the symplectic structure in the Darboux form  with some new
variables, $P$ and $Q$, say, that contained the canonical
variables of [DNG] and [GB] systems, and thus, we will  see the
contribution to quantum level of the [GB] term when we add it 
the  [DNG] action in string theory, however,  this we will discuss in future works (partial results  is given in  \cite{17b}).  \\
\newline
\noindent \textbf{Acknowledgements}\\[1ex]
This work  was  supported by CONACyT under grant 44974-F .  The author wants to thank R. Capovilla for the support and  friendship that he has offered me. The author also thanks the referee for drawing my attention on Ref. \cite{5b,5c,5d}.  \\
\newline
\newline
\newline
\newline
\noindent \textbf{Appendix}\\[1ex]
\newline
In this section, we will try to stablish a conection with  the results given in \cite{5d} taking a particular case of  the results found in this paper.
For our  purposes,  we will consider a relativistic extended object $\Sigma$, of dimension $p$, embedded in arbitrary fixed $(N+1)$- dimensional background spacetime $\{M, g_{\mu \nu}\}$, $\Sigma $  is described locally by the spacelike embedding $x^\mu=X^\mu (u^A)$, where $x^\mu $ are local coordinates for the background spacetime, $u^A$ local coordinates for $\Sigma $, and $X^\mu $
 the embedding functions $(\mu, \nu, ..., = 0,1,..., N$, and $A,B,...,=1,...,p$).\\
 The tangent vectors to $\Sigma$ are defined by $\epsilon_A^{{\mu}}= \partial_A X^\mu \partial_\mu$, so that the  positive definite metric  on $\Sigma $ is 
\begin{equation}
h_{AB}=g(\epsilon_A, \epsilon_B)= \epsilon_{A}^{{\mu}} \epsilon_{B}^{{\nu}} g_{\mu \nu },
\end{equation}
note that we can construct out of the metric $h_{AB}$ the intrinsic geometry of $\Sigma $ (for more details about the intrinsic and extrinsic geometry see \cite{5, 7}).\\
We consider now the time evolution of $\Sigma $ in spacetime.  We denote its trajectory, or worldvolume, by $m$. It is a oriented time like surface in spacetime. Now the shape functions become time-independent, $X^\mu= X^\mu(\tau, u^A)$, where $\tau$ is a coordinate that labels the leafs of the foliation of $m$ by $\Sigma s$.\\
The time evolution of the embedding functions for $\Sigma$ in to the worldvolume can be written as
\begin{equation}
\dot{X^\mu}= N \eta^{\mu} + N^A \epsilon_{A}^{\mu},
\end{equation}  
where $\eta^\mu$ is the unit (future-pointing) timelike normal to $\Sigma$ into $m$, $N$ is called the lapse functions and $N^A$ the shift vector \cite{5aa}.
We can note that the content of this equation is simply that the time evolution of $\Sigma$ is into the worldvolume $m$.  \\ The geometry of the worldvolume can be represented in parametric form by the embedding functions $x^\mu= \chi ^\mu (\xi ^a)$, where $\xi^a=(\tau, u^A)$ are local coordinates for $m$ (see the paragraph below Eq.(3)), and $\chi^\mu$ the embedding functions ($a= 0,1,...,p$).\\ 
The  tangent vectors to $m$, $e_a= e_a ^{\mu} \partial_\mu$, decompose in a part tangential to $\Sigma$ and part along the time evolution of $\Sigma$, 
\begin{equation}
e_a ^{\mu}= \left(
\begin{array}{rr}
\dot{X^{\mu}}   \\
\epsilon_A^{\mu}  \\
\end{array}
\right).
\end{equation}
Thus, the worldvolume induced metric, $\gamma_{ab}= e_a ^{\mu}e_b ^{\nu} g_{\mu  \nu} $\cite{5,7}, takes the familiar ADM form 
\begin{equation}
\gamma_{ab}= \left( 
\begin{array}{cc}
-(N^2 - N^AN^B)h_AB &h_{AB}N^B \\
h_{AB}N^B& h_{AB}
\end{array}
\right) \,.
\end{equation}
we can see that the worldvolume element is given by $\sqrt{- \gamma}= N \sqrt{h}$. The expression given in (31) is the same  found in \cite{5aa, 5d}.\\
With these results, we can calculate the total momentum $P^\mu(\Sigma )$ of the spatial hypersurfaces, this is, taking in account the eqution (14) we find 
\begin{equation}
 P^\mu(\Sigma )= \int_{\Sigma} \tau_a P^{a \mu}= - \alpha \int_{\Sigma} \sqrt{h} \eta^{\mu},
\end{equation} 
where $\eta^{\mu}= \tau^a e_a^{\mu}$ is the unit velocity vector at a given point on $\Sigma$. The momentum  density $P^{a \mu}$  is not only tangent to the worldsheet, it also lies parallel to the tangent vector, $e_a^{\mu}$ .
In this sense, the extremal surfaces, like geodesics, are self-parallel.
It is important to mention, that the momentum (32) is found in \cite{5d}, however, in \cite{5d} the geometric content  of (32) is confuse. Thus,  with the results of this paper  is  easier  to obtain it and interpret in a  geometry context.\\
On the other hand, using (31) we can write the [DNG] action for bosonic p-branes as $S=  \int d \tau L[X, \dot{X}] $, with the lagrangian given by 
\begin{equation}
L[X, \dot{X}]= - \alpha\int_{\Sigma} d^pu\sqrt{- \gamma} = - \alpha \int_{\Sigma}d^pu N \sqrt{h} ,
\end{equation}
 thus, using (32) and (33) we find that 
 \begin{equation}
 H[X, P]= \int d^pu [P_{\mu} \dot{X}^\mu] - L[X, \dot{X}]=0,
 \end{equation}
 such as expected from worldvolume reparametrization invariance the hamiltonian vanish. In this manner, we can see that the hamiltonian is a linear combination of the phase space constraints that generate reparametrizations  $\{F_0, F_A \}$. This constraints, using (32) is geven by 
 \begin{eqnarray}
 F_{A}&=& P_{\mu} \epsilon_{A}^{\mu}=0, \\
 F_{0}&=& P^2+ \mu^2 h=0, 
  \end{eqnarray}
  and the Hamiltonian takes the form  
  \begin{equation}
  H[X,P]= \int_{\Sigma_{\tau}} [\lambda F + \lambda^A F_{A}],
  \end{equation}
  where $\lambda$ and $\lambda^A$ are Lagrange multipliers enforcing the constraints. To find this Lagrange multipliers we calculate the first Hamilton equation, using the Eqs. (35), (36) and (37) we find  
  \begin{equation}
  \dot{X}^\mu = \frac{\delta H}{\delta P_\mu}= 2 \lambda P^\mu+ \lambda^A \epsilon_{A}^{\mu},
  \end{equation}
  from Eqs. (29) and (38) we can identify that; $\lambda= \frac{N}{2 \alpha \sqrt{h}}$ and $\lambda^A=N^A$. We can see that all this results was obtained in \cite{5aa, 5d} by a  different way.\\
  To finish, is important to mention that the constraint (35) generates diffeomorphisms tangential to $\Sigma$; $F_0$ is a universal constraint for all reparametrizations invariant action, they are the generators of worldvolume reparametrizations. All this symmetries,  in the gauge invariance of $\omega$ (Eq. (1)) are included;  in \cite{5} we prove that $\omega$ is invariant  under spacetime diffeomorphisms and worldvolume  reparametrizations. This is, our simplectic structure   $\omega$ inherits the covariant properties of the deformation formalism, from which it has been constructed \cite{5}.  \\
 In addition to this work, we know that Polyakov action  of the p-branes is more suitable for quantization, however, with the results of this paper we will see in coming  works if  there are any progress with [DNG] action,  the details of this work with the Polyakov action we will discuss in the future. The reason is  that we need  to develop a  covariant formalism of deformation as in \cite{7} and \cite{34}  for the Polyakov action, and to do  an analisis like  this work for such geometry. On the other hand,  important results in this direction can be found  in \cite{5c, 5d}.
\newline
\\[1em]

\end {document}